>

# Aerial Push-Button with Two-Stage Tactile Feedback using Reflected Airborne Ultrasound Focus

Hiroya Sugawara, Masaya Takasaki, Keisuke Hasegawa

***Abstract*—We developed a new aerial push-button with tactile feedback using focused airborne ultrasound. This study has two significant novelties compared to past related studies: 1) ultrasound emitters are equipped behind the user's finger and reflected ultrasound emission that is focused just above the solid plane placed under the finger presents tactile feedback to a finger pad, and 2) tactile feedback is presented at two stages during pressing motion; at the time of pushing the button and withdrawing the finger from it. The former has a significant advantage in apparatus implementation in that the input surface of the device can be composed of a generic thin plane including touch panels, potentially capable of presenting input touch feedback only when the user touches objects on the screen. We experimentally found that the two-stage tactile presentation is much more effective in strengthening perceived tactile stimulation and feeling of input completion when compared with a conventional single-stage method. This study proposes a composition of an aerial push-button in much more practical use than ever. The proposed system composition is expected to be one of the simplest frameworks in the airborne ultrasound tactile interface.**

## I. INTRODUCTION

### *1.1. Current midair airborne ultrasound haptics and its application to aerial input interfaces*

AMONG the principles of tactile presentation, a method using intense midair focused ultrasound generated via airborne ultrasound phased arrays has many basic and application studies that so far have been conducted [1][2] since the first proposal [3][4]. This method produces acoustic radiation force on the skin surface that is in contact with no tactile devices. This nonlinear acoustic phenomenon caused by focused ultrasound emitted from the phased array can present spatially pinpoint tactile stimulus to any position of an unspecified number of people's body surface, which is epochal in the realm of tactile display devices.

Early research of midair haptics attempted the presentation of vibrotactile stimuli mainly for the palms using amplitude modulation of driving waveform of transducers [5][6]. Then, methods to present the shape of an object by using specifically designed ultrasonic wavefronts were researched [7][8][9]. In addition, spatiotemporal extension of the modulation method of ultrasonic focus has been widely studied in an attempt to accomplish even stronger tactile stimulations or a greater variety of tactile experiences [10-14]. Regarding different haptic modalities, presenting cool [15] or warm [16] sensations were conducted. Thus, the research of tactile presentation by airborne ultrasound has been showing its continuous growth.

As applied scenarios, studies that offer experiences of touching objects having texture and shape are often found [17][18][19]. As another application category, many aerial tactile interfaces have been developed focusing on the importance of the intuitive feeling of input offered via tactile feedback. Among them, the studies of tactile input interface inside a car that the driver uses while driving have been widely conducted [20][21][22]. The most common midair tactile input interface is push-buttons with haptic feedback, which has many precedent examples [23][24][25]. The tactile input interfaces integrated with the visual contents were also proposed, where the system overlapped airborne ultrasound tactile presentation on the aerial footage [26], or non-contact visible push-buttons are realized over the acoustic semi-transparent layer under which the phased arrays are installed [27].

### *1.2. Remaining issues in preceding aerial tactile input systems*

As reviewed above, there are a lot of studies in input interface using airborne ultrasound haptics. On the other hand, some technological aspects that can be further investigated are listed as follows:

1) Realization of an input surface requiring no specific devices under the user's hand
2) Pursuit of aerial tactile feedback that provides users with feelings of input completion with good fidelity

Regarding the first aspect, tactile sensations presented by ultrasound focusing use direct waves in most cases. In those situations, since most areas in the workspace are occupied by ultrasound transducers, which prevent any other devices from being placed there, considerable space is needed for the workspace composition. Also, for focusing ultrasound to create a focal point with acoustic energy sufficiently localized to be felt as a pinpoint tactile spot, the focal positions cannot be placed just above the transducers but at several hundreds of millimeters away from them with commonly used 40 kHz transducers. Therefore, a large depth ahead of the users' hand is needed, possibly preventing ultrasound tactile presentation from being introduced in actual living areas.

Next, regarding the second aspect, the realization of tactile feedback offering users the feeling of physically interacting with buttons or levers has not necessarily been conducted in previous related studies. In other words, current midair haptic interfaces indeed allow the users to distinguish whether their input process is completed or not in a tactile manner. However, there is still room for improvement in creating the input feeling with a more convincing intuitiveness. Since the behaviors of physical input devices in the real world are familiar to us, an possible approach to realize such aerial input feelings is the tactile feedback imitating these behaviors of the devices. This approach is often adopted in the study of tactile devices with physical contact with users' bodies. With airborne ultrasound tactile presentation, it is essentially difficult to compose rigid mechanical components since the finger motion cannot be constrained. Nevertheless, there is a possibility to present sufficiently convincing input feeling if the skin sensation can be engendered with the essential aspects of physical reactions of corresponding input devices being conveyed to the users' perception.

### *1.3. Aims and scientific contributions in this paper*

For the above problems in midair ultrasound tactile presentation, this study proposes a new method for the construction of the aerial push-button via the following approaches:

- The virtual input surface is set just above the ultrasound-reflecting plane and the ultrasound focus for tactile stimulation is formed on the user's fingertip surface when impinging the input surface. We use airborne ultrasound phased arrays placed apart from the input surface with their emission planes facing downward, which emit focused ultrasound from behind the user's finger via the acoustic reflecting surface.
- The user's input finger motion is detected by an optical sensor. The tactile stimulation is then presented to the fingertip at specific two times: when the finger pushes into and is withdrawn from the virtual button.

The former is novel in the device composition in midair ultrasound tactile presentation and has great benefits in that any electronic devices do not need to be placed under the hand and the reflecting plate can be used for other purposes when the aerial button is not activated. Even if the user's hand is dirty or wet, the tactile stimuli can be presented without damaging the device as the conventional system setup would. Moreover, it is another great advantage that natural visuo-tactile feedback can be realized by replacing the reflection surface with visual displays like LCDs, which will offer a visible push-button with three-dimensional tactile feedback presented only when the user's finger interacts with it.

Regarding the latter, there is a preceding study about ultrasound tactile presentation via two stages [25], which supposes the construction of a virtual push-button with a height of 100 mm presenting to the palm. In that study, tactile feedback is given at two phases where the user's finger touches the button surface and completes the pushing motion. Contrastingly, our study supposes the user's finger motion within a little millimeter in depth, and the tactile stimulation imitating the reaction force from the internal mechanism of the button is presented, at the times of finger insertion and withdrawal. Much literature says that actual mechanical buttons have a mechanism where the reaction force from the button varies suddenly at specific finger positions during input motion [28][29][30]. We aim to imitate such impulsive tactile responses given to the user's finger during relatively small finger displacement. Although the static reaction force from actual push-buttons cannot be expressed in the airborne ultrasound tactile presentation, we expect that producing input feeling is possible to some degree by only presenting the aforementioned impulsive tactile feedback.

This paper shows the method of constructing a new airborne ultrasound tactile interface, which is designed based on the above strategies. We demonstrate the superiority of the above method of a two-stage tactile presentation in terms of the presented tactile input feeling of a push-button and subjective strength of tactile stimulation through the psychophysical experiment.

## II. Principle of the Proposed System and Strategy of Tactile Presentation

### *2.1. Acoustic radiation force by focused ultrasound by the airborne ultrasound phased array*

The ultrasound phased array (Fig. 1), which is used as an airborne tactile presentation device [31], has an ultrasound emission surface on which many air-coupled ultrasound transducers are arranged in a two-dimensional lattice pattern. This device can focus ultrasound on any position on the human skin surface (this position is hereinafter referred to as the 'focal point') by appropriately setting phase delays on each vibrator driven at the same frequency. Here, when the ultrasound is blocked off by any solid objects, acoustic radiation force is exerted on the object surfaces parallel to the normal of the surfaces, with its intensity depending on the amplitude of the sound. Therefore, by placing the finger near the focal point, the finger surface is pressed by acoustic radiation force being localized and can perceive tactile stimulation in the air. Via this tactile stimulation by the ultrasound focal point, we do not feel the ultrasound waveform itself as vibration. Instead, the envelope of the waveform is what we feel as vibrotactile stimulation [14]. Although the radiation force on the finger generated by ultrasound phased arrays is as small as tens of millinewtons, it can present sufficiently perceivable tactile stimulation by modulating the ultrasound focus temporally or spatially.

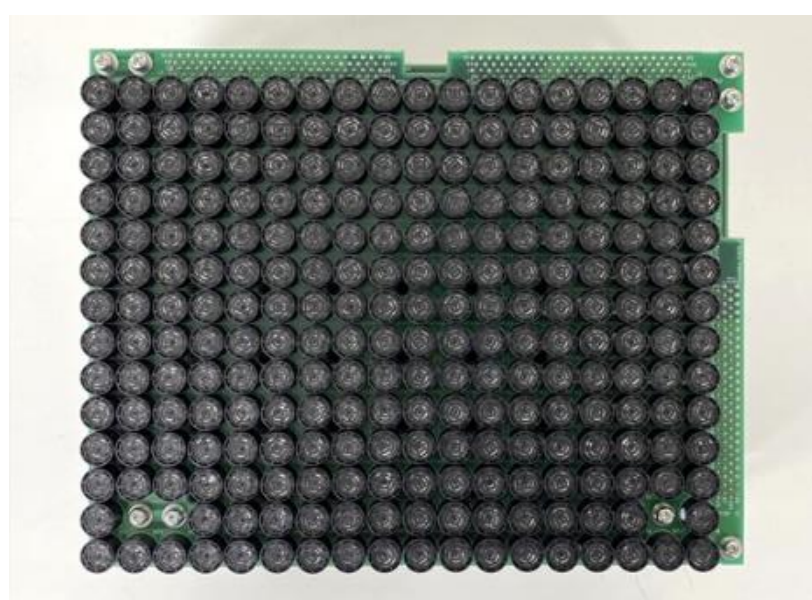

Fig. 1. The 40 kHz-driven ultrasound phased array used in the experiments

*2.2. Formation of tactile stimulation point by focusing reflected wave emitted from behind the finger*

Based on the principle of airborne ultrasound tactile presentation stated above, we propose a push-button system having airborne tactile feedback as shown in Fig. 2. Unlike the related studies, we form the resultant ultrasonic focal point above the ultrasound-reflecting surface that does not have any function of tactile presentation. Any solid materials that are sufficiently smooth and flat can serve as this surface.

In the proposed system, the phased arrays are placed above the input surface and emit ultrasound so that the original focal point is set at the back side of the reflecting surface. On this occasion, a corresponding mirrored focal point is generated at a position of a mirrored image of the original focus with respect to the reflecting surface. Here, we conduct the phase control of transducers to achieve the ultrasound focusing resulting at this mirrored focal point. The phased arrays are placed so that the radiation surfaces tilt by 45 degrees from the normal direction of the reflecting surface and the ultrasound is radiated from diagonal behind the finger. This expects that the radiation force around the focal point presses the finger pads upwards, although some portion of emitted ultrasound may be occluded by the user's hand.

*2.3. Implementation of two-stage tactile presentation*

As mentioned in the introduction, we aim to provide convincing tactile experiences of physically completing the input motion of a virtual push-button. To this end, instantaneous tactile stimulation by a burst wave of a short term is presented when the finger pushes into the button (hereinafter referred to as the "downward" finger motion) and after that, the finger leaves the button (hereinafter referred to as the "upward" finger motion), respectively (Fig. 3). Tactile presentation by burst wave is reported to be a suitable method to present "click feeling" in prior research [32].

In the two-stage tactile presentation, it is necessary to detect the motion of the user's fingertip. In this study, we equipped a pair of an IR LED and a phototransistor on the left and right side of the input surface (Fig. 4). This paired electronic components is connected to the microcontroller circuit, which detects the finger insertion and withdrawal on the input surface by detecting the IR ray getting into the phototransistor being decreased/increased to a certain amount due to the motion of lowering/raising the finger near the input surface.

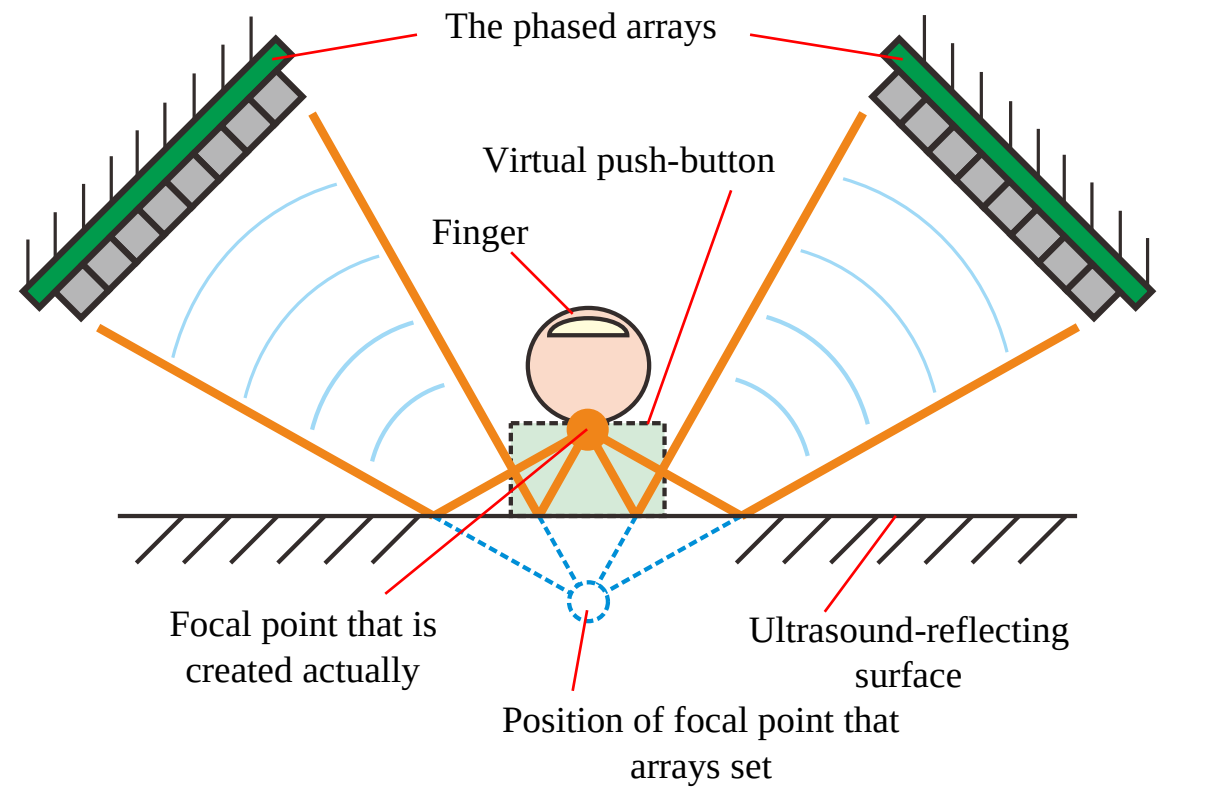

**Fig. 2.** Proposed method of constructing a midair push-button with tactile feedback

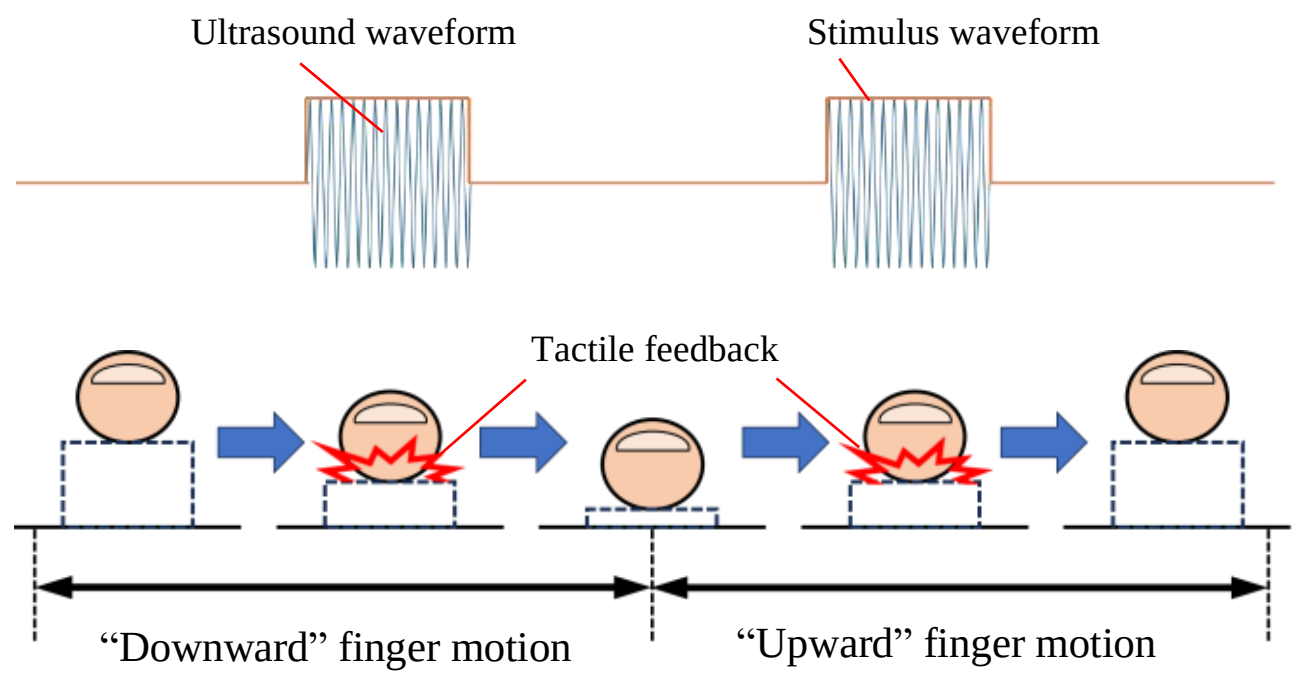

**Fig. 3.** The timing of presenting ultrasonic tactile feedback (upper row) during the finger motion (lower row) in our system

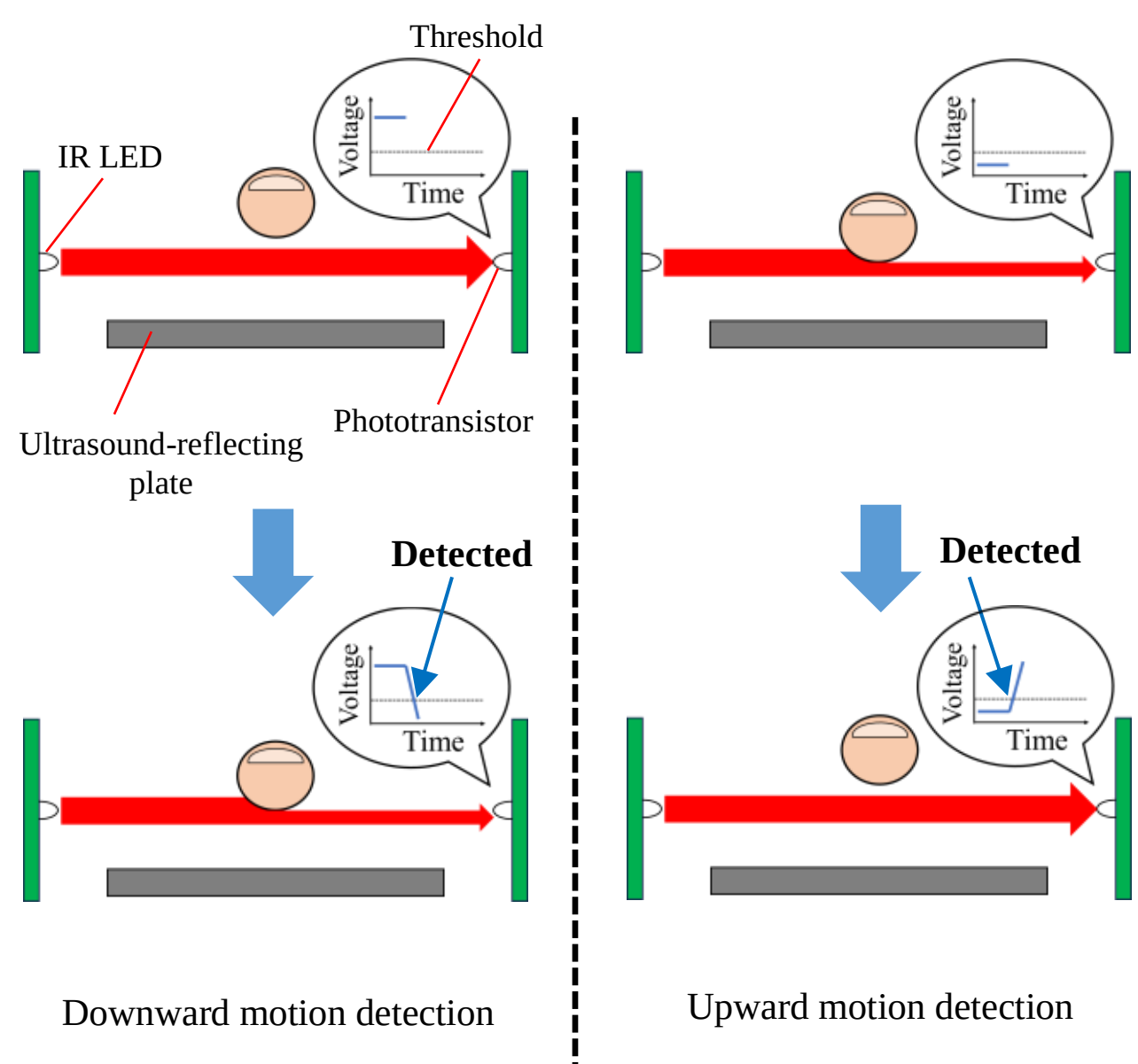

**Fig. 4.** Schematic description of the finger movement detection during two-stage tactile presenting using IR LED and phototransistor.

## III. Tactile Display System and Procedures of the Experiments

### *3.1. Fabricated prototype system of the aerial push-button*

We fabricated a prototype device of the aerial push-button and evaluated the effectiveness of the tactile presentation based on the aforementioned principles as a midair tactile input interface. Fig. 5 shows the device and its schematic view. We placed a reflecting plate of ultrasound on the workspace center and placed a pair of an IR LED and a phototransistor on the left and right side to detect the finger input. They were placed in a vertical movable stage that allowed the distance to be adjusted between the reflecting plate and the pair of the LED and the phototransistor. As the reflecting plate a 120 mm × 120 mm × 10 mm aluminum plate was used, and the LED and the phototransistor were placed at an interval of approximately 200 mm. In front of the LED, a 3 mm-thickness metal plate with a drilled 1 mm-diameter hole was placed to enhance the directivity of the LED ray. Two 40 kHz-driven phased arrays (AUTD3 [31]) were placed so that their center was located at the position 200 mm above the reflecting surface. The distance between the centers of ultrasound emission planes of the arrays was 270 mm. A microcontroller unit (MCU) connected to a PC for driving the phased arrays monitored the voltage output of the phototransistor to detect finger insertion and withdrawal. The ultrasound focus was immediately formed at the moment that the MCU detected that the voltage signal crossed the predetermined threshold. Regarding the threshold of voltage, we set the threshold for "upward" motion detection higher than that for "downward" to suppress chattering.

### *3.2. Basic characteristics of the prototype system*

For an input interface, the time delay from the user's input motion to the presented sensory feedback is an important practical aspect. Since the input in this system is detected by the LED ray getting into the phototransistor being attenuated due to the finger movement, we observed the voltage waveform of the phototransistor and the ultrasound waveform to evaluate this time delay. We confirmed that the duration from the voltage reaching the threshold to the start of ultrasound emission was less than 15 ms. This delay, considering that the standard frame rate of current video games is 60 fps, is equivalent to 1 frame and is not expected to undermine the presented input feeling.

Next, we measured the force near the focal point by reflection wave to evaluate the magnitude of the radiation force on the finger during the system operation. As shown in Fig. 6, we equipped a 10 mm diameter round bar on the jig placed on the electronic scale and put it just above the reflecting plate. In this situation, we set the distance of 1 to 10 mm between the round bar and the reflecting plate and formed the focal point at a height of 1 to 10 mm from the reflecting plate, both with an interval of 1 mm. In this experimental setting, the weight of the jig measured by the electronic scale decreased by the upward radiation force exerted on the round bar and thus the magnitude of the force can be measured.

Fig. 7 shows the relationship between the distance between the round bar and the plate and the strength of force. Note that the resolution of the electronic scale was 0.1 g. This indicates that the strength of force periodically varied with a period of 4 to 5 mm with respect to the variation of the distance between the round bar and the plate. A possible explanation for this is that the ultrasound wavelength was approximately 8.5 mm and may have caused standing waves that reflected between the round bar and the plate, whose strength periodically varied depending on the bar-plate distance.

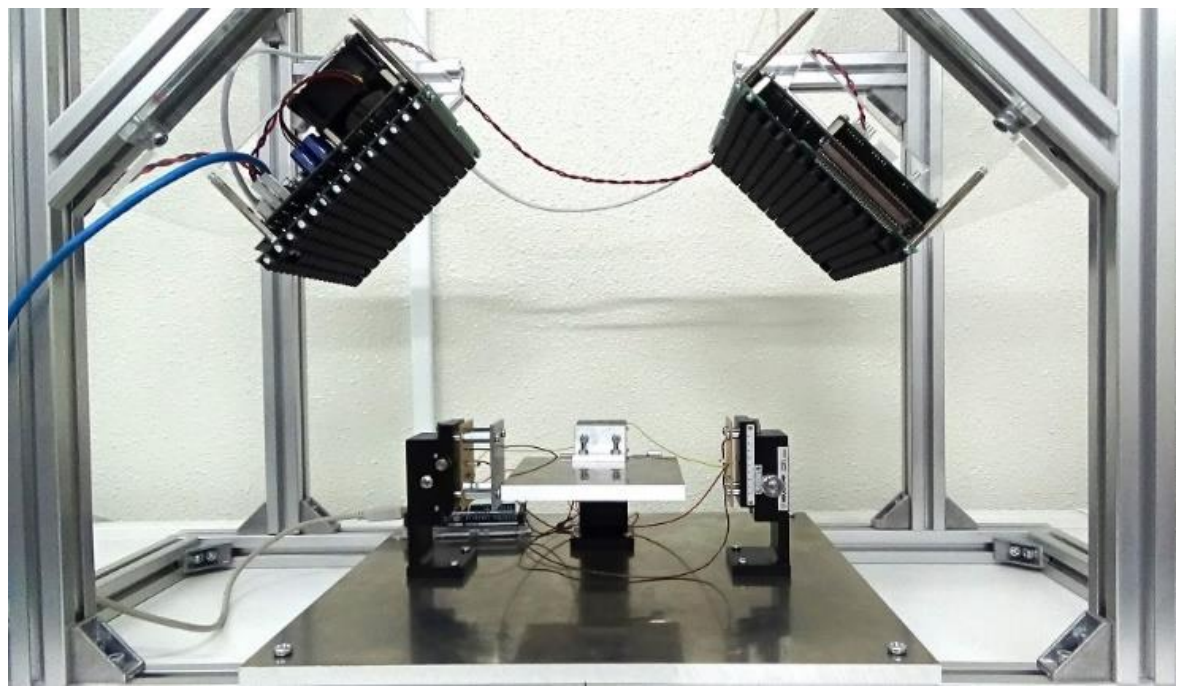

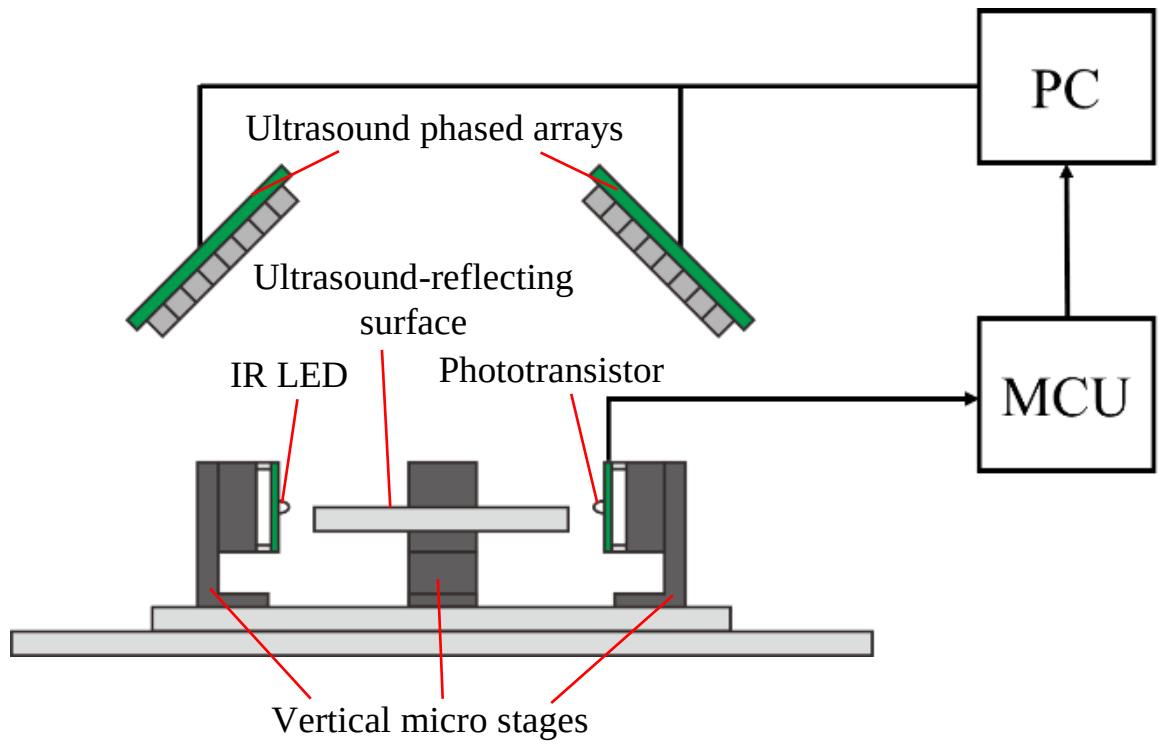

**Fig. 5.** Experimental apparatus and its schematic view

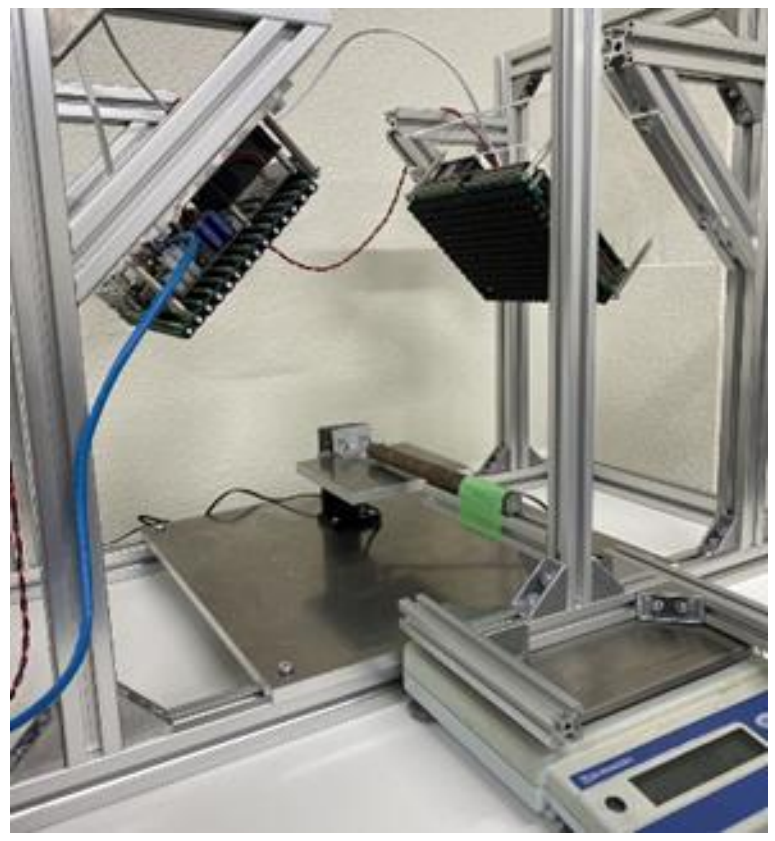

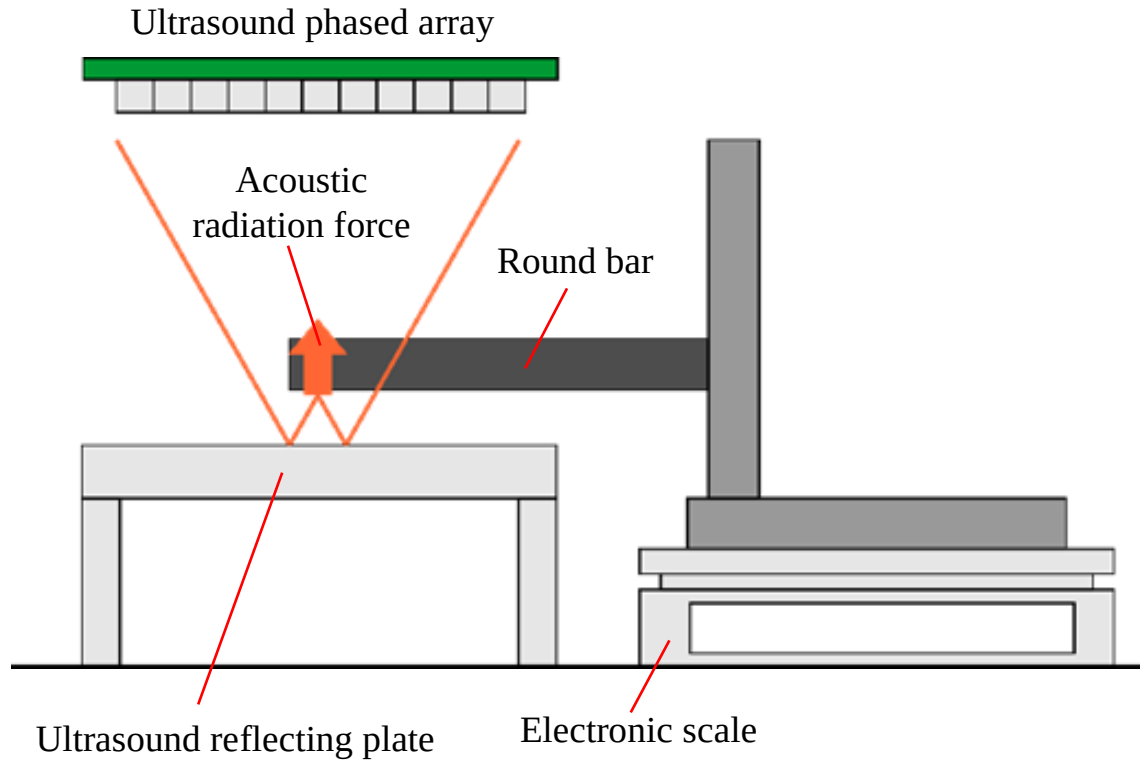

**Fig. 6.** The method of measuring acoustic radiation force

Considering this result and the stroke of a general key or button, we tried the two-stage tactile presentation in our finger while changing the height of the LED and phototransistor from the reflecting plate between 3 to 5 mm. According to our subjective evaluation, the input feeling of the button was the most prominent at 3 mm. Therefore, the height of the LED and the phototransistor pair and the focal point from the reflecting plate were set to 3 mm and we decided to present tactile stimulation in this setting.

### *3.3. experimental procedures of tactile feedback presentation*

We conducted psychophysical experiments to assess the effectiveness of tactile presentation during “downward” and/or “upward” finger motions on improvement of input feeling and sensation strength. The experiments were approved by the ethical committee of Saitama University (No. R5-E-25). The experimental procedures are as follows.

10 people in their early twenties participated, with their ages ranging between 20 and 23. In the experiments, as shown in Fig. 8, the following three conditions were set: (I) tactile presentation during only “downward” motion, (II) tactile presentation during only “upward” motion, and (III) tactile presentation during both “downward” and “upward”. For each condition, there were two cases of burst times: 50 ms and 100 ms. Furthermore, we set another two conditions, where the finger touched the reflecting plate and it did not. Therefore, there are 12 conditions in total.

The experimental procedures are described below.

1. The participants sat on a chair as shown in Fig. 9, and listened to white noise through headphones so that they could not hear the driving sound of the phased arrays.
2. At first, they placed the index finger of their right hand on the reflecting plate. When they heard a beep lasting one second through the headphones, they lifted their finger off the plate and then moved it up and down within the vertical range of approximately 10 mm while touching the plate. One of the six conditions (with fingers touching the plate) was presented in a random order during this time. Note that participants were only informed of the basic principle of the device and not informed of the method of tactile presentation in experimental conditions or under which conditions the system was being activated.
3. After five seconds, they stopped the finger when heard a 0.5 second beep and evaluated “Stimulus strength” and “Feeling of input” on a seven-point Likert scale.

The tactile presentation was conducted five times for each of the above six conditions and thus this procedure was repeated 30 times.

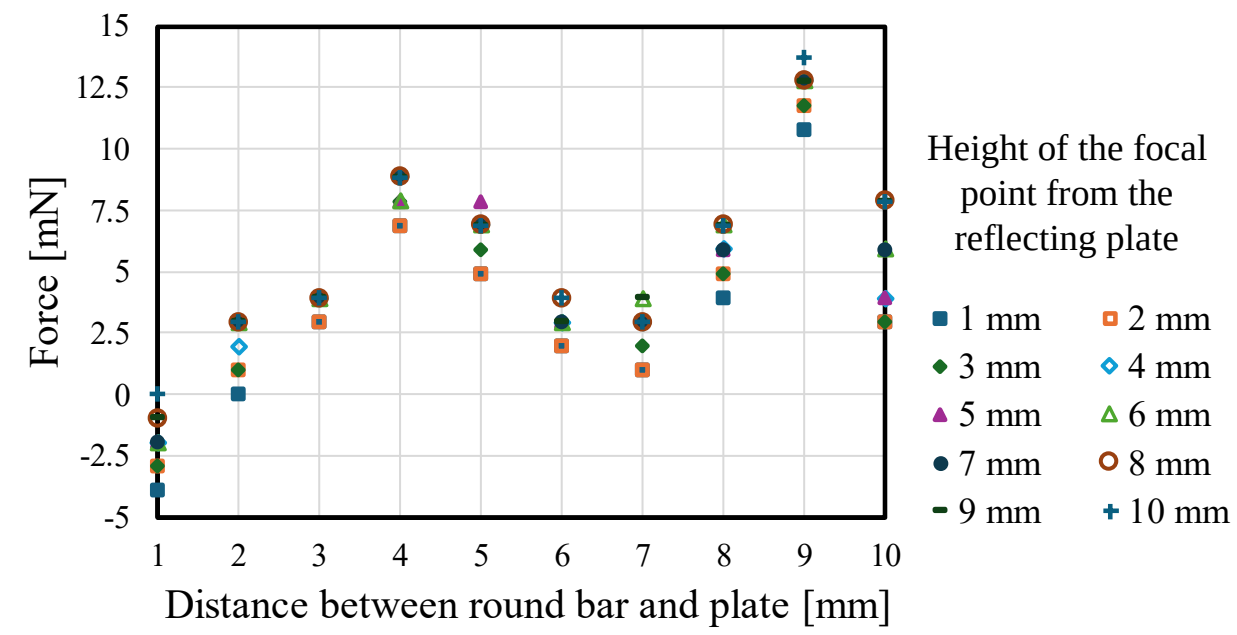

**Fig. 7.** The relationship between the radiation force, distance between round bar and the reflecting plate, and the focal height

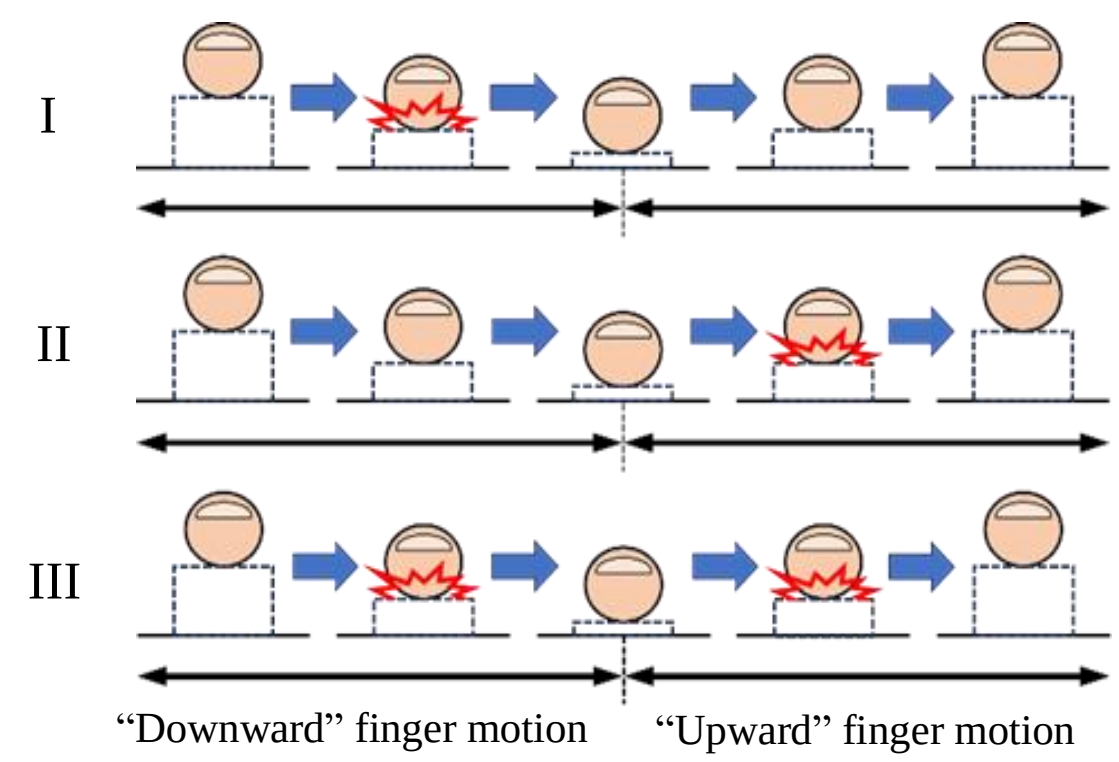

**Fig. 8.** Three types of presenting tactile feedback in the experiments.

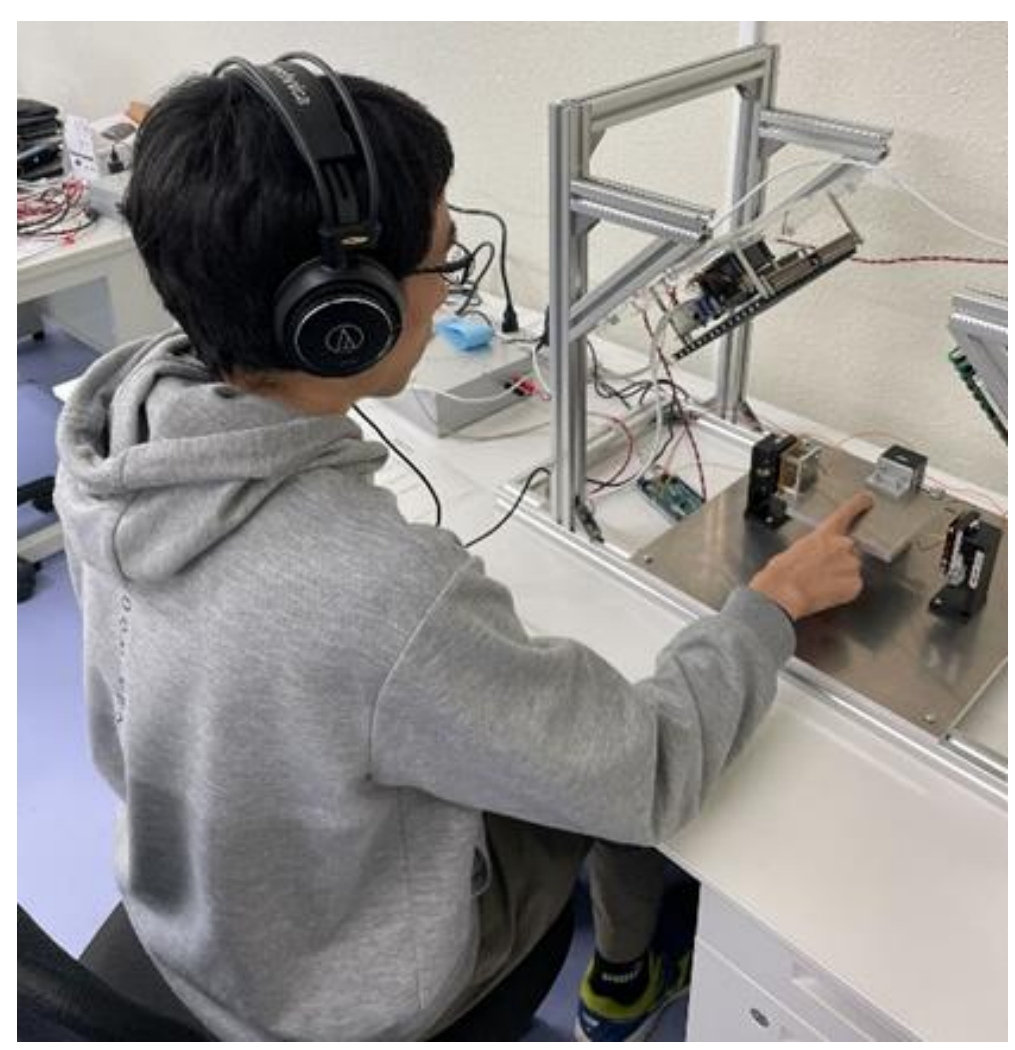

**Fig. 9.** Picture of the experimental situation

Next, the participants repeated the same procedures (1. to 3.) another 30 times without touching the plate and completed the same evaluations, which corresponded to the other remaining six conditions.

Prior to the actual experimental trials, participants repeated this procedure three times as a training session, with touching and non-touching conditions.

## IV. Experimental Results

### 4.1. Stimulus strength

The upper graph in Fig. 10 shows the box plot of the participant's ratings about the subjective stimulation strength for each condition with participants' fingers touching the plate (resulting in 50 total scores per condition). The lower graph in Fig. 10 shows the corresponding averages and standard errors for each condition. First, we performed the Shapiro-Wilk test for the data sets to assess their statistical normality, which was denied under all conditions (note that the data normality was also denied under the other six conditions where the participants' fingers were not touching the plate). Hence, we conducted Friedman's test and confirmed the significant difference between the three conditions in the same burst time. We calculated the p-values for those conditions using Bonferroni's multiple comparison test. We used statistical analysis software EZR [33]. Note that symbols * and ** in the figures denote significant differences at p-values smaller than 0.05 and 0.01, respectively.

In the cases with users' fingers touching the plate, significant differences in stimulus strength were found for (I)-(II) and (I)-(III) at a p-value smaller than 0.01 when the burst time was 50 ms. When the burst time was 100 ms, a significant difference was observed for (II)-(III) at a p-value smaller than 0.05 and for (I)-(II) and (I)-(III) at a p-value smaller than 0.01.

As shown in Fig. 11, in the cases with users' fingers not touching the plate, significant differences were found for (I)-(II) at a p-value smaller than 0.05 and for (II)-(III) at a p-value smaller than 0.01 both when the burst time was 50 ms and 100 ms. The results show that tactile presentation in "upward" motion is perceived more strongly than that in only "downward" motion when it is displayed alone and it is combined with "downward" stimulation. On the other hand, the combination of "downward" and "upward" stimulation was not significantly stronger than "upward" stimulation alone in most cases.

### 4.2. Feeling of input

Next, Figs. 12 (touching the plate) and 13 (not touching the plate) show the box plot of the participant's ratings for each condition (50 scores per condition) and the averages and standard errors in the feeling of input. For the assessment of significant differences among conditions, we calculated the p-value in the same procedure.

In the case of users' fingers touching the plate, significant differences were found for (I)-(II) at a p-value smaller than 0.05, and for (I)-(III) at a p-value smaller than 0.01 when the burst time was 50 ms. When the burst time was 100 ms, the significant differences were confirmed for (I)-(II) and (II)-(III) at a p-value smaller than 0.01.

In the case of users' fingers not touching the plate, significant differences were found for (II)-(III) at a p-value smaller than 0.05 and for (I)-(III) at a p-value smaller than 0.01 when the burst time was 50 ms. When the burst time was 100 ms, significant differences were confirmed for (I)-(III) and (II)-(III) at a significant level of 0.01. These show that the tactile presentation with the combination of both “downward” and “upward” is significantly more effective than that with “downward” alone for the cases of both touching and non-touching the plate.

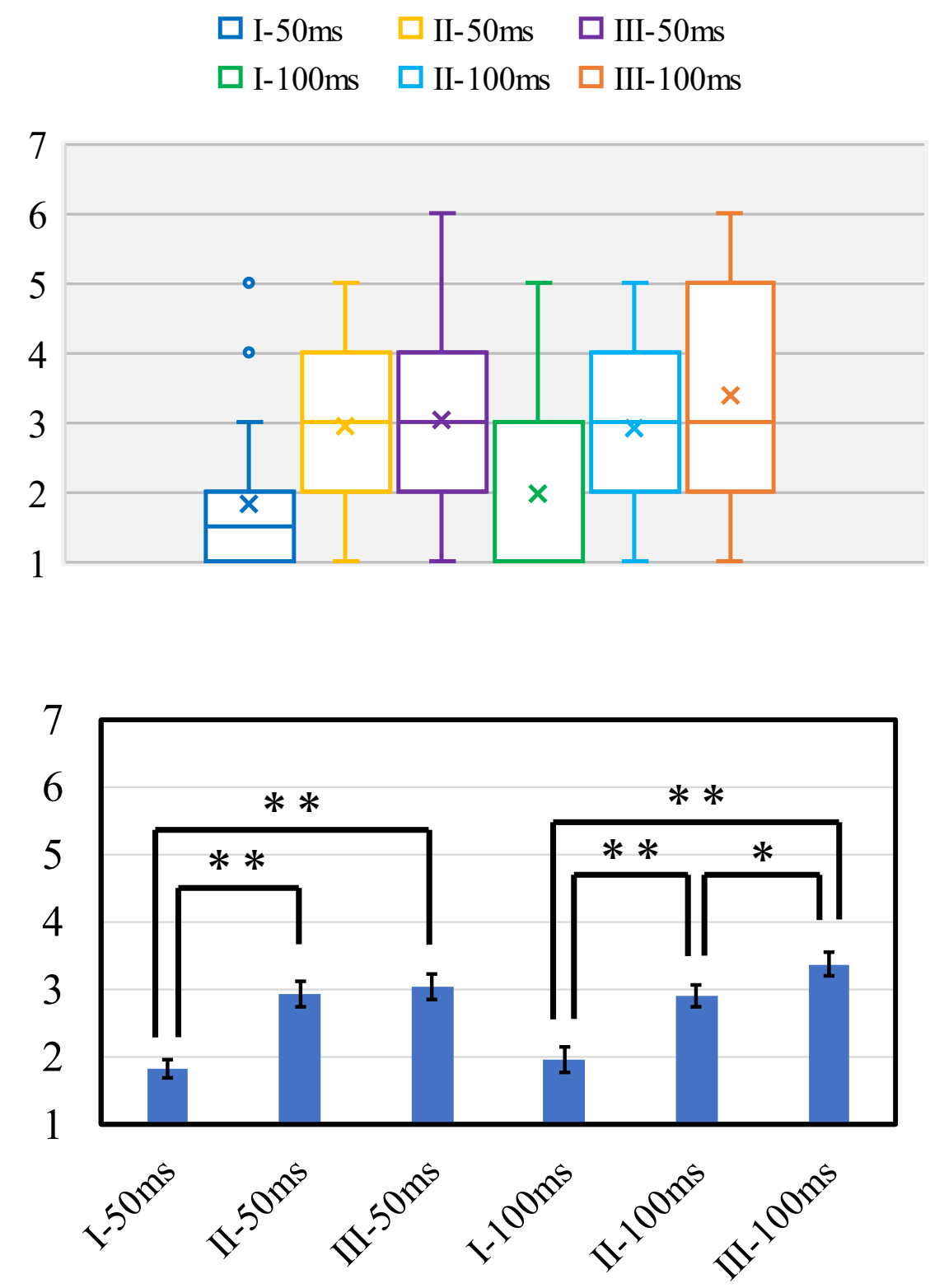

**Fig. 10.** Experimental results of stimulus strength with fingers touching the plate

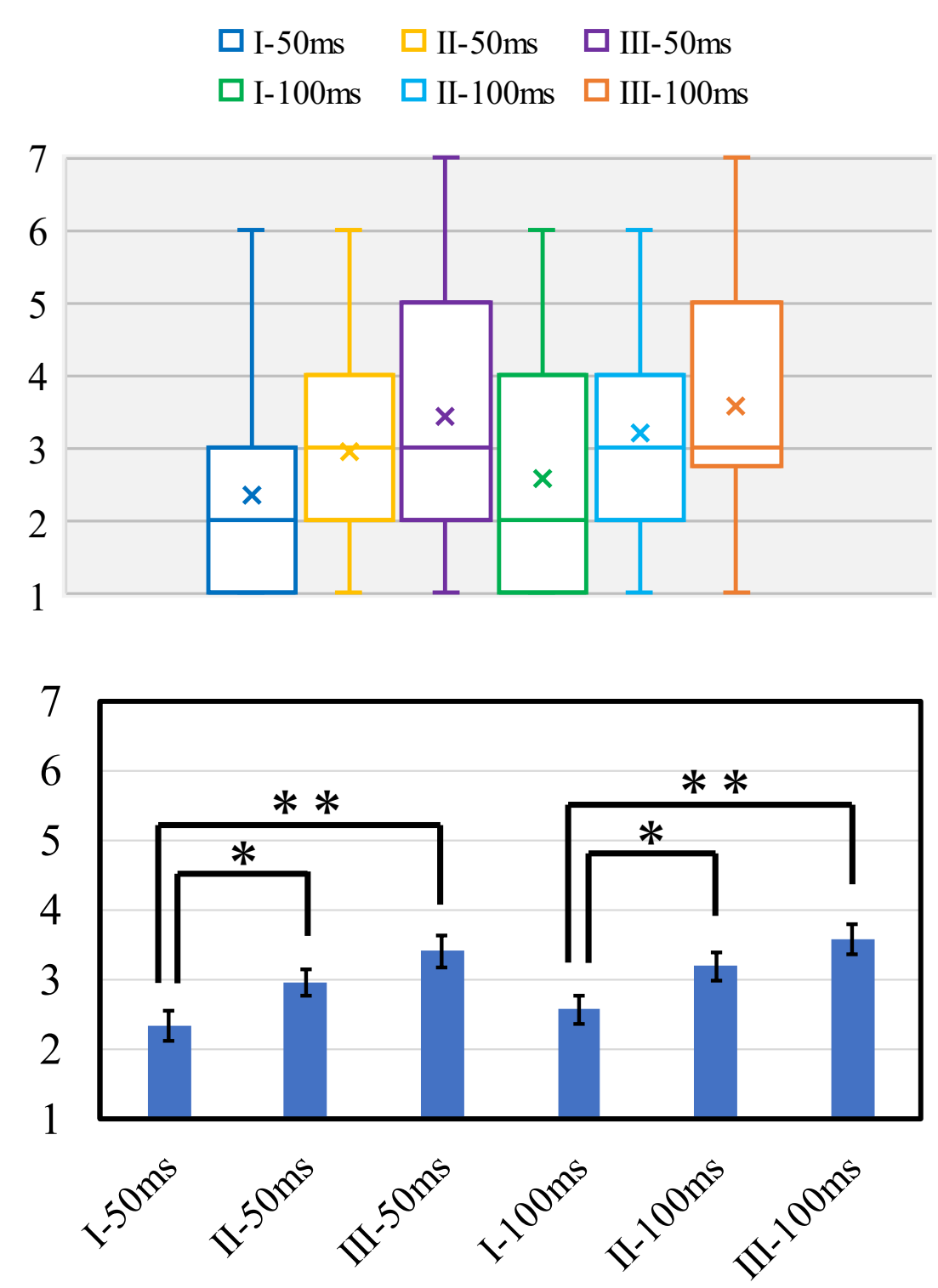

**Fig. 11.** Experimental results of stimulus strength with fingers kept away from the plate

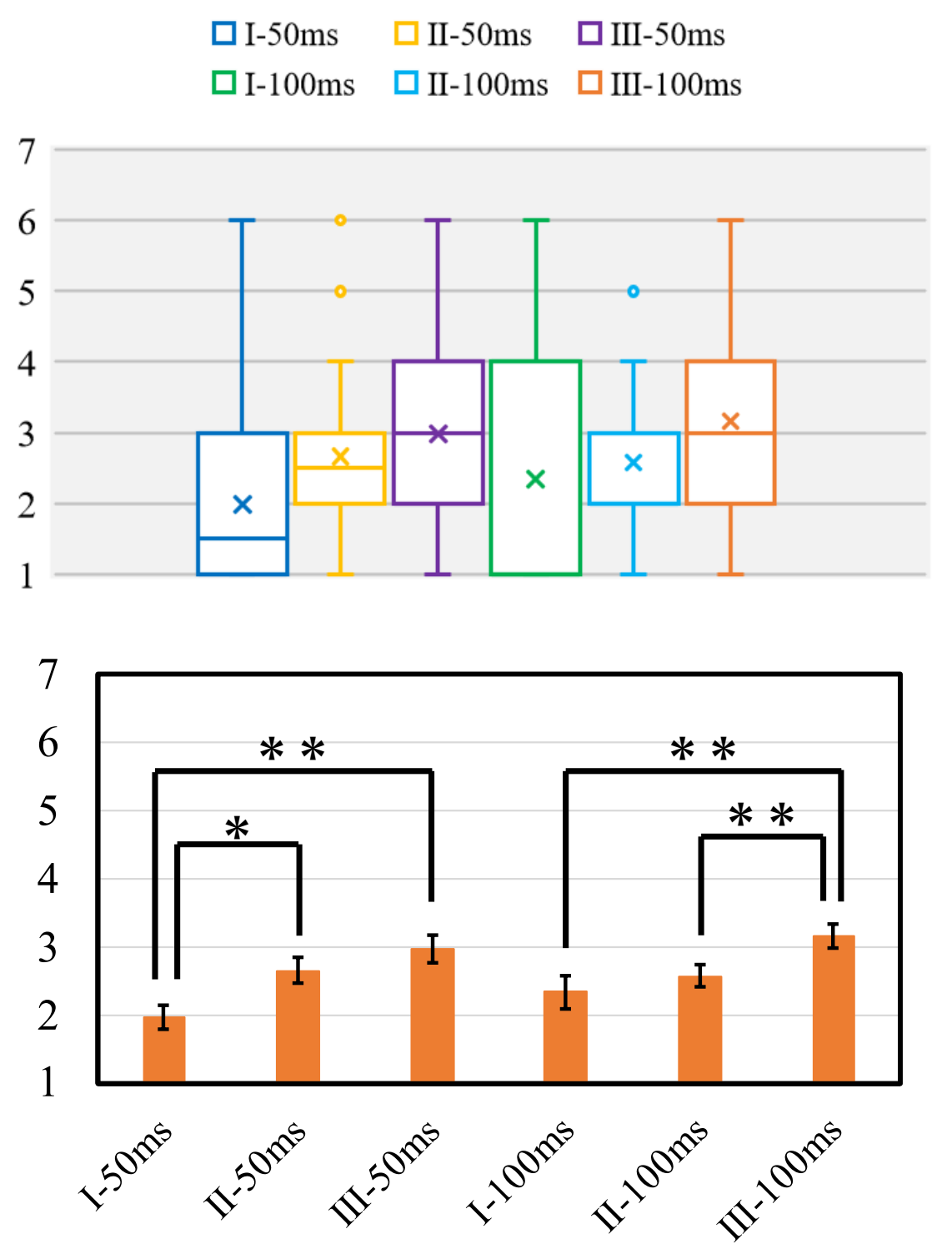

**Fig. 12.** Experimental results of feeling of input with fingers touching the plate

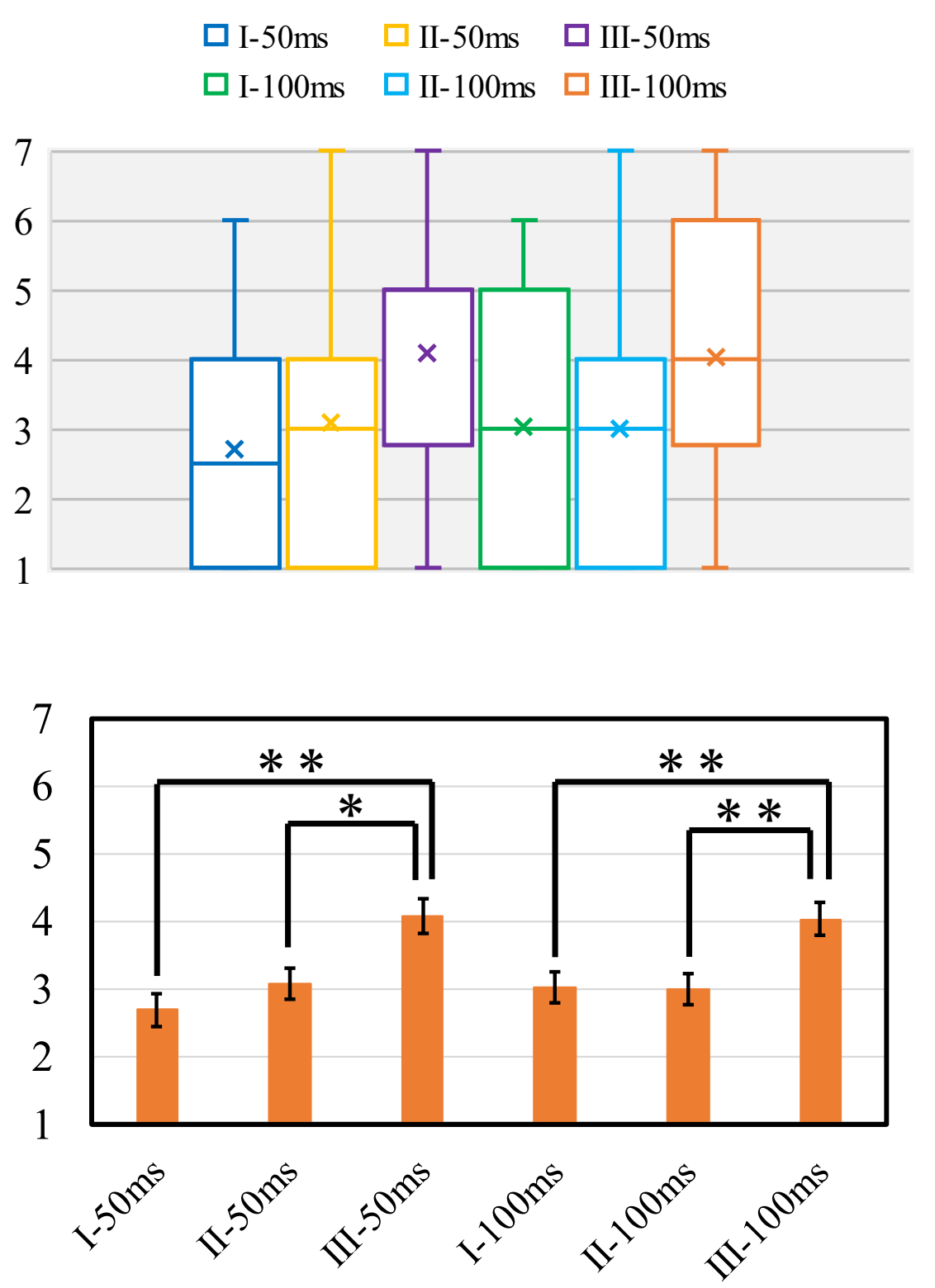

**Fig. 13.** Experimental results of feeling of input with fingers kept away from the plate

Concerning the overall results that feedback during the “upward” motion was perceived more strongly than “downward”, we calculated the p-values by Wilcoxon signed-rank test to confirm the difference between touching and non-touching in the same condition. The results are shown in Figs. 14 and 15. These figures show that the cases with “downward” alone showed significant differences in perceived stimulation strength between touching and non-touching conditions. On the other hand, the cases with “downward” alone and that combined with the “upward” ones showed significant differences in perceived feeling of input completion between touching and non-touching conditions. No significant difference was found with respect to the difference in the burst time between any conditions.

## V. Discussions

The experimental result that the stimulation in “upward” motion was perceived more strongly than that in “downward” may be explained by an assumption that the subjective strength of perceived tactile stimulation on the skin surface is different between during finger flexion and extension. This hypothesis is based on the fact the moment at feedback during “upward” motion is when the finger muscles are anatomically relaxed. This is the state that an upward force works at the same time in the direction that expedites muscle relaxation. The situation is completely opposite at the moment at the “downward” case in that the muscles are contracted then.

In respect of stimulus strength, feedback during the “upward” motion alone was significantly stronger than that during the “downward” motion alone, whereas no significant difference between “upward” alone and the combination of “downward” and “upward” was confirmed. At the same time, the combination of both “downward” and “upward” significantly enhanced the feelings of input completion, compared with “downward” alone or “upward” alone. Thus, in the setting of our experiments, simply strengthening the subjectively perceived tactile stimulation does not necessarily lead to an improvement of the feeling of input completion, which shows a possibility that the two-stage tactile presentation during input motion leads to realistic expression of physical responses of a push-button.

The distance between the ultrasound-reflecting plate and the input surface was small in our experimental setup, which allows the user to rest their finger by placing it on the plate during the input motion. This is similar to the situation that occasionally takes place, where the user does not operate the device but places the finger on it when using a push-button or a keyboard. Our experiments show that when the user was allowed to do this, the perceived stimulation magnitude at the tactile presentation was maintained during the two-stage feedback compared to that in the situation where the user held the finger in the air. On the other hand, the feeling of input completion was significantly different during the two-stage tactile feedback depending on whether the fingers touched the surface or not. This is interpreted as that the tactile and haptic sensation caused by touching the actual plate affected the feeling of input.

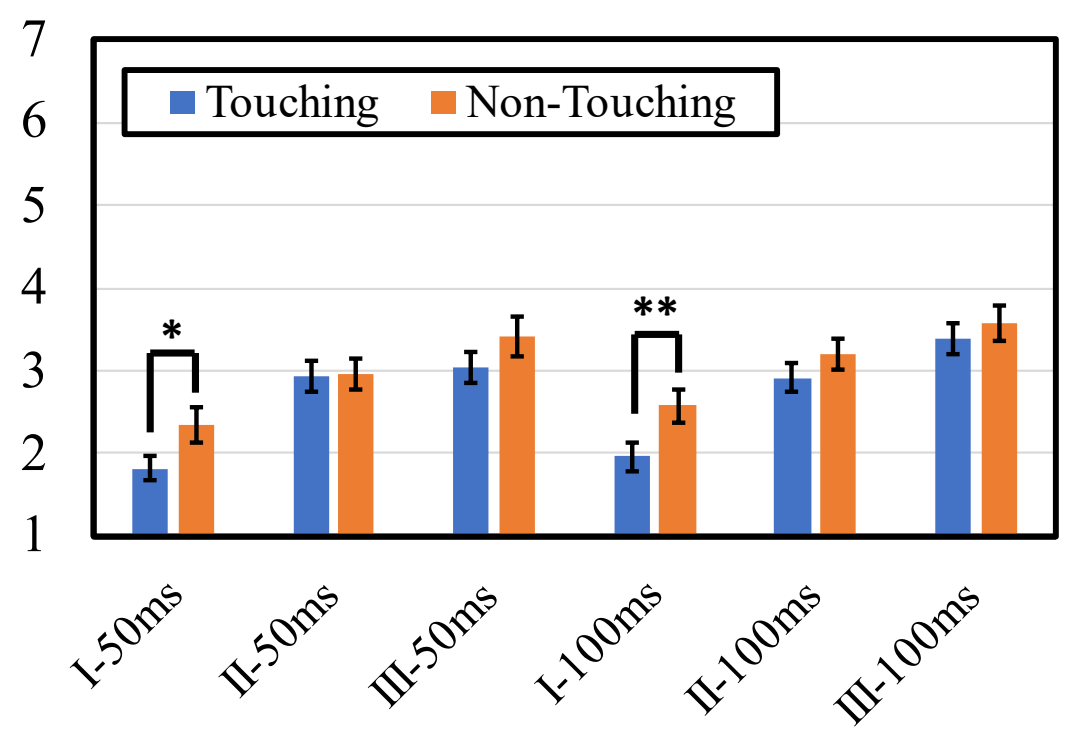

**Fig. 14.** Comparison of perceived stimulation strength between touching and non-touching

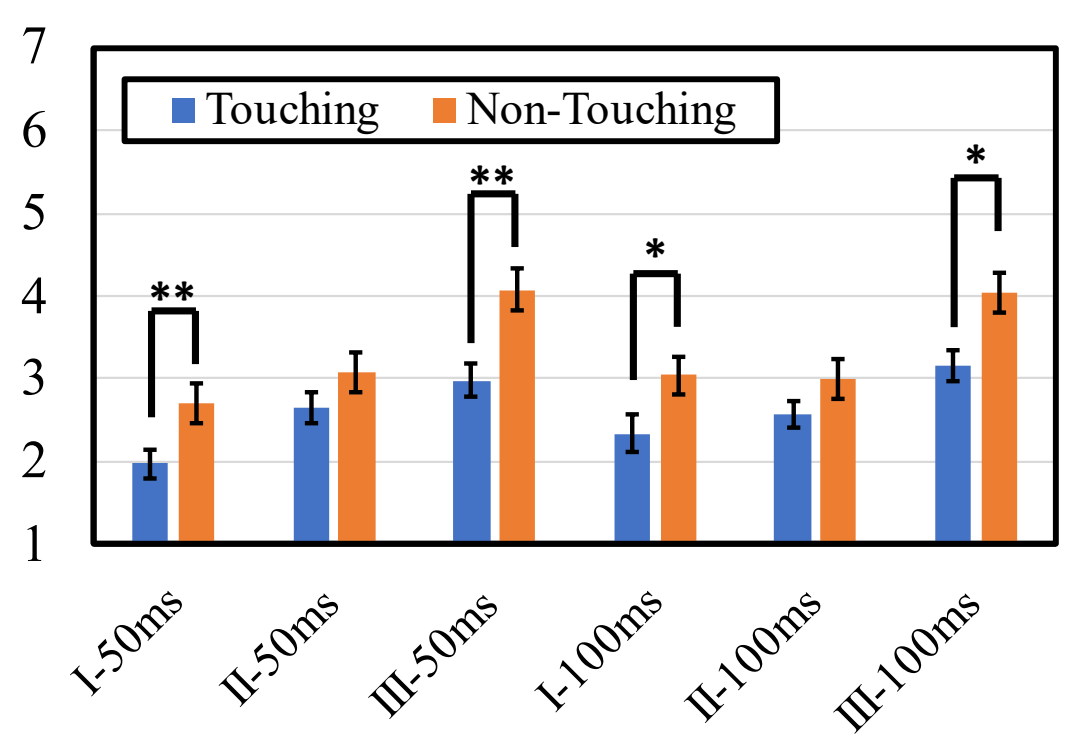

**Fig. 15.** Comparison of perceived feeling of input between touching and non-touching

## VI. CONCLUSIONS

In this study, we propose a virtual aerial push-button with a midair input surface that exerts upward acoustic radiation force on the finger just above the ultrasound-reflecting plate. This push-button has two-stage tactile feedback presented during the user's "downward" and "upward" finger motion, which was proven to be effective for the improvement of the feeling of input compared to that presented during either "downward" or "upward" motion alone. Also, we clarified that this is not necessarily relevant to the variation of subjective strength of tactile stimulation caused by the change in the tactile feedback timings.

It is expected that these results will be widely applied as the design principle of presented stimulation in airborne ultrasound tactile interfaces. Also, the device composition that we proposed in this study is effective in terms of hygiene and maintenance readiness since the ultrasound phased arrays are placed above the operation area and it reduces the possibility that the user touches or pollutes the arrays. Furthermore, the way users operate the interface can be flexible with our system configuration since the user can hold the finger in the air or place it on the ultrasound-reflecting plate.

As mentioned above, it is possible that the acoustic reflecting plate can be constructed using various types of video display devices. Therefore, a visuo-tactile interface with the appearance of a keyboard or a button can be implemented, which provides users with an intuitive visible aerial operation surface by the achievement of this study, without any special visual display component. At this stage, when the finger contacts the touch panel directly during the operation, another problem of how to integrate the haptic sensation by the solid touch surface with the tactile stimulation presented by airborne ultrasound will emerge. In addition, the extension of this achievement toward the realization of another input interface different from a push-button is one of the practical directions of future studies.